\documentclass[12pt]{article}

\advance\voffset by -2.0cm \advance\hoffset by -1.25cm
\usepackage{amsmath}
\usepackage{amssymb}
\usepackage{amsthm}
\usepackage{amsfonts}

\numberwithin{equation}{section}
\newtheorem{theorem}{Theorem}[section]

\theoremstyle{definition}

\newcommand{\CC}{\mathbb{C}} 
\newcommand{\ZZ}{\mathbb{Z}} 
\newcommand{\PP}{\mathbb{P}} 

\DeclareMathOperator{\Sym}{Sym} 

\newcommand{\M}{{\mathcal M}} 
\newcommand{\be}{\begin{equation}}
\newcommand{\ee}{\end{equation}}

\newlength{\oldcolsep}

\begin{document}

\title{Extending the Belavin-Knizhnik ``wonderful formula" by the characterization of the Jacobian}
\author{Marco Matone}\date{}

\maketitle

\begin{center} Dipartimento di Fisica ``G. Galilei'' and Istituto
Nazionale di Fisica Nucleare \\
Universit\`a di Padova, Via Marzolo, 8 -- 35131 Padova,
Italy\end{center}

\bigskip

\begin{abstract}
A long-standing question in string theory is to find the explicit
expression of the bosonic measure, a crucial issue also in
determining the superstring measure. Such a measure was known up to
genus three. Belavin and Knizhnik conjectured an expression for
genus four which has been proved in the framework of the recently
introduced vector-valued Teichm\"uller modular forms. It turns out
that for $g\geq4$ the bosonic measure is expressed in terms of such
forms. In particular, the genus four Belavin-Knizhnik ``wonderful
formula" has a remarkable extension to arbitrary genus whose
structure is deeply related to the characterization of the Jacobian
locus. Furthermore, it turns out that the bosonic string measure has
an elegant geometrical interpretation as generating the quadrics in
$\PP^{g-1}$ characterizing the Riemann surface. All this leads to
identify forms on the Siegel upper half-space that, if certain
conditions related to the characterization of the Jacobian are
satisfied, express the bosonic measure as a multiresidue in the
Siegel upper half-space. We also suggest that it may exist a super
analog on the super Siegel half-space.

\end{abstract}

\newpage

\section{Introduction}

Denote by $y_1,\ldots,y_{3g-3}$ some complex analytic coordinates on
the moduli space of genus $g\geq2$ compact Riemann surfaces  $\M_g$,
and by $\tau$ the Riemann period matrix. According to the
Belavin-Knizhnik theorem the genus $g$ partition function of the
Polyakov bosonic string \cite{Polyakov:1981rd} is
\cite{Belavin:1986cy}
$$
Z_g=\int_{\M_g}{F\wedge \bar F\over (\det {\rm Im}\tau)^{13}} \ ,
$$
where
$$
F:=F(y_1,\ldots,y_{3g-3})dy_1\wedge\cdots\wedge dy_{3g-3} \ ,
$$
is a holomorphic $(3g-3,0)$ form which vanishes nowhere on $\M_g$. Furthermore, $F$ has a second-order pole, due to the tachyon, at the Deligne-Mumford boundary of $\M_g$.

Consider the Siegel
upper half-space
$${\frak H}_g:=\{Z\in M_g({\Bbb C})\mid {}^tZ=Z,\mathop{\rm Im} Z>0\} \ ,$$
and the Thetanullwerte
$$
\chi_k(Z):=\prod_{\delta\hbox{ even}} \theta[\delta](0,Z) \ ,
$$
$Z\in{\frak H}_g$, with $k=2^{g-2}(2^g+1)$.
For $g=2$ it has been shown
in \cite{BelavinTV,MorozovDA,DHokerQP} that $F$ is proportional to
$${\wedge_{i\leq j}^2d\tau_{ij} \over
\chi_{5}^2(\tau)} \ .
$$
For $g=3$, it has been conjectured
in \cite{BelavinTV,MorozovDA} and proved in
\cite{IchikawaTI,DHokerCE}, that $F$ is proportional to
$${\wedge_{i\leq j}^3d\tau_{ij}\over \chi_{18}^{1/2}(\tau)} \ .
$$
Finding the higher genus expression of $F$ is a long-standing
problem involving basic questions which are not only of a purely
technical nature, rather they concern the foundations of string
theories. In particular, it turns out that the problem of the
effective characterization of the Jacobian locus (Schottky problem)
is deeply related to string theory. This is already evident once one
considers that the Fay]'s trisecant formula \cite{Fay}, which is the
higher genus version of the bosonization (Cauchy determinantal)
formula, characterizes, even if not effectively, the Jacobian.
Furthermore, the even unimodular lattices $E_8$ and $D_{16}^+$ of
the heterotic strings turn out to characterize the Jacobian at $g=4$
(se below), and perhaps all trigonal curves or $n$-gonal curves, for
some $n\equiv n(g)$.

In string theory the Schottky problem arises from the beginning,
that is for $g=4$. In this case, Belavin-Knizhnik
\cite{Belavin:1986cy} and Morozov \cite{MorozovDA} conjectured that
\be F={d\tau_{11}\wedge \cdots \wedge\widehat
{d\tau}_{ij}\wedge\cdots \wedge d\tau_{44}\over S_{4ij}(\tau)} \ ,
\label{wonderful}\ee where
$$S_{4ij}(Z):={1+\delta_{ij}\over 2}{\partial F_4(Z)\over \partial Z_{ij}} \ ,$$
with
$$F_g:=2^g
\sum_{\delta\hbox{
even}}\theta^{16}[\delta](0,Z)-\bigl(\sum_{\delta\hbox{
even}}\theta^{8}[\delta](0,Z)\bigr)^2 \ .
$$
It turns out that $F_4$, the Schottky-Igusa form, vanishes only on the Jacobian. Furthermore, there is a nice relation
between $F_g$ and the theta series $\Theta_\Lambda$
corresponding to $\Lambda=E_8$ and
$\Lambda=D_{16}^+$
$$
F_g=2^{-2g}(\Theta_{D_{16}^+}-\Theta_{E_8}^2) \ .
$$
Eq.(\ref{wonderful}) has several consequences. First, it shows that the bosonic string is strictly related to the question of characterizing the Jacobian. Furthermore,
it follows that the genus four bosonic partition function can be expressed as residue formula in the Siegel upper half-space \cite{Belavin:1986cy}\cite{MorozovDA}
\be
Z_4=\int_{{\frak H}_4} {1\over  (\det {\rm Im} Z)^{13}}\Bigg|{\wedge_{i\leq j}^4dZ_{ij}\over F_4(Z)}\Bigg|^2 \ .
\label{greatwonderful}\ee
As stressed by Morozov in \cite{Morozov:2008xd}, such a ``wonderful formula" has no attracted the due attention.
Its elegance and simplicity suggest the intriguing possibility that there exists a formulation of the bosonic string on the Siegel upper half-space.
This would be of considerable interest also in superstring theory as one may investigate its formulation on the super analog of the Siegel upper half-space.

A more rigorous derivation of Eq.(\ref{wonderful}) has been proposed
in \cite{Guilarte}. More recently, Eq.(\ref{wonderful}) has been
proven in \cite{Matone:2011ic} in the context of vector-valued
Teichm\"uller modular forms, a framework that shed light on its
geometrical origin. The aim of the present paper is to further
investigate its structure and then generalize it to the case of
$g\geq5$. In section 2 we shortly review the vector-valued
Teichm\"uller modular forms, in particular it turns out that for
$g\geq4$ the bosonic measure is expressed in terms of them. In
Section 3 we discuss how such relation implies that the bosonic
string measure characterizes the Riemann surface in $\PP^{g-1}$.

In section 4 we show that the vector-valued Teichm\"uller modular
forms are expressed as the determinant of the coefficients of the
polynomials of degree $n$ in $\PP^{g-1}$, this is the key point that
will lead, as a corollary, to the higher genus extension of the
Belavin-Knizhink formula (\ref{wonderful}), derived in section 5. We
will also consider a natural candidate for the extension of
(\ref{greatwonderful}) and will suggest it should be possible
finding its super analog on the super Siegel half-space.

\section{The bosonic measure and vector-valued Teichm\"uller modular form}

In this section we shortly review the main results of
\cite{Matone:2011ic}. In particular, we will see some that for $g\geq4$
the bosonic measure is expressed in terms of the vector-valued
Teichm\"uller modular forms.

Denote by $C$ a compact Riemann surface of genus $g\geq2$ and
by
$$M_n:={g+n-1\choose n} \ ,$$
and
$$N_n:=(2n-1)(g-1)+\delta_{n1} \ , $$
the dimension of ${\rm Sym}^n\, H^0(K_C)$ and $H^0(K_C^n)$ respectively.
Set
$$K_n:=M_n-N_n \ . $$
We will also use the notation
$$
M\equiv M_2=g(g+1)/2 \ , \qquad N\equiv N_2=3g-3 \ , \qquad K\equiv K_2=(g-2)(g-3)/2 \ .
$$
Let $\{\alpha_1,\ldots,\alpha_g,\beta_1,\ldots,\beta_g\}$ be a symplectic basis of $H_1(C,\ZZ)$. Denote by
$\{\omega_i\}_{1\le i\le g}$ the basis of $H^0(K_C)$ with the standard normalization $\oint_{\alpha_i}\omega_j=\delta_{ij}$ and by
$\tau_{ij}=\oint_{\beta_i}\omega_j$ the Riemann period matrix.

Consider the moduli space of principally polarized abelian varieties
${\cal A}_g={\frak H}_g/{\rm Sp}(2g,\ZZ)$. According to Torelli's theorem, the morphism
$$
i :\M_g \rightarrow {\cal A}_g \ ,
$$
which on (geometric) points takes the algebraic curves to its Jacobian, is injective. The question
of characterizing the image of $i$ is the Schottky problem.

Denote by $\mathcal{C}_g \stackrel{\pi}{\longrightarrow}{\M}_g$ the
universal curve over  $\M_g$ and by $L_n=R\pi_*(K^n_{\mathcal{C}_g/\mathcal{M}_g})$ the vector bundle on $\mathcal{M}_g$ of rank
$N_n$ with fiber $H^0(K_C^n)$ at the point of $\mathcal{M}_g$ representing $C$.
Let $\lambda_n:=\det L_n$ be the determinant line bundle. The
Mumford isomorphism is \cite{Mumford}
$$
\lambda_n\cong\lambda_1^{\otimes c_n}\ ,
$$
where $c_n:=6n^2-6n+1$. The Mumford forms $\mu_{g,n}$ are the unique,
up to a constant, holomorphic section of
$\lambda_n\otimes\lambda_1^{-\otimes c_n}$ nowhere vanishing on
$\M_g$. Let $\{\phi^n_i\}_{1\le i\le N_n}$ be a basis of
$H^0(K_C^n)$, $n\geq2$. It turns out that
$$
\mu_{g,n}={\kappa[\omega]^{(2n-1)^2}\over
\kappa[\phi^n]}{\phi^n_1\wedge\cdots\wedge\phi^n_{N_n}\over
(\omega_1\wedge\cdots\wedge\omega_g)^{c_n}} \ ,
$$
where
$$\kappa[\omega]:=
{\det\omega_i(p_j)\sigma(y)\prod_1^gE(y,p_i)\over
\theta\bigl(\sum_{1}^gp_i-y-\Delta\bigr)\prod_1^g\sigma(p_i)
\prod_{i<j}^gE(p_i,p_j) }\ ,$$
and
$$
\kappa[\phi^n]:={\det \phi_i^{n}(p_j)\over
\theta\bigl(\sum_{1}^{N_n}
p_i-(2n-1)\Delta\bigr)\prod_{1}^{N_n}\sigma(p_i)^{2n-1}\prod_{i<j}^{N_n}
E(p_i,p_j)}\ ,
$$
$n\geq2$, where we followed the notation of \cite{Matone:2011ic}. Note that $\mu_{g,n}$ is independent of the choice of the basis of $H^0(K^n_C)$.

In the case of $n=2$ the Mumford form is strictly related to the Belavin-Knizhnik form $F$,
that they obtained from the
curvature form of the determinant of Laplace operators,
$$
F={\kappa[\omega]^{9}\over
\kappa[\phi^2]}\phi^2_1\wedge\cdots\wedge\phi^2_{3g-3} \ .
$$
The fact that the bosonic string measure is determined by the Mumford form $\mu_{g,2}$ has been first discovered by Manin \cite{Manin:1986gx}
who noticed that $c_2=13$ in Mumford's formula coincides
with the half of the string critical dimension.
The expression of $\mu_{g,2}$
 in terms of
theta functions
has been given by Beilinson and Manin \cite{BeilinsonZW} whereas $\mu_{g,n}$
has been obtained
by Verlinde and Verlinde \cite{VerlindeKW} and Fay \cite{FayMAM}.

It is useful to use a single indexing, for example, instead of $\omega_i\omega_j$, with $i,j=1,\ldots,g$, we may use the shorter notation $\omega^{(2)}_i$, $i=1,\ldots,g(g+1)/2$. More
generally, let us consider the basis
$\tilde\omega_1^{(n)},\ldots,\tilde\omega_{M_n}^{(n)}$ of $\Sym^n
H^0(K_C)$ whose elements are symmetrized tensor
products of $n$-tuples of vectors of the basis
$\omega_1,\ldots,\omega_g$, taken with respect to an arbitrary
ordering. The image
$\omega_i^{(n)}$, $i=1,\ldots, M_n$, of $\tilde\omega_i^{(n)}$ under
$\psi:\Sym^n H^0(K_C)\to H^0(K_C^n)$ is surjective for $g=2$
and for $C$ non-hyperelliptic of genus $g>2$.

Let us consider $\mu_{g,2}$. For $g>3$ one has $K>0$, so that using $N$ elements of
${\rm Sym}^2 H^0(K_C)$ as basis for $H^0(K_C^2)$ leads to free indices.
More generally for $\mu_{g,n}$ this happens when $K_n>0$ and one uses $N_n$ elements of ${\rm Sym}^n H^0(K_C)$
as basis for $H^0(K_C^n)$.
This leads to the concept of vector-valued Teichm\"uller modular forms \cite{Matone:2011ic}.
For example, in the case $n=2$, $g>3$, one is led to consider wedge products $\wedge_{k=1}^{3g-3} \omega_{i_k}\omega_{j_k}$, or,
upon taking the Kodaira-Spencer map
$$\omega_i\omega_j\mapsto {1\over2\pi i}d\tau_{ij} \ ,$$
the wedge products
$$
\wedge_{k=1}^{3g-3} d\tau_{i_kj_k} \ ,
$$
so that there are free indices.

The vector-valued Teichm\"uller modular forms are \cite{Matone:2011ic}
\be [i_{N_n+1},\ldots,i_{M_n}|\tau]:=\epsilon_{i_1,\ldots,i_{M_n}}{\kappa[\omega^{(n)}_{i_1},\ldots,\omega^{(n)}_{i_{N_n}}]\over\kappa[\omega]^{(2n-1)^2}} \ ,
\label{uarmo}\ee
$i_1,\ldots,i_{M_n}\in\{1,\ldots,M_{n}\}$. Note that
\be
[i_{N_n+1},\ldots,i_{M_n}|\tau]=\epsilon_{i_1,\ldots,i_{M_n}}{\omega^{(n)}_{i_1}\wedge\cdots
\wedge\omega^{(n)}_{i_{N_n}}\over (\omega_{1}\wedge\cdots
\wedge\omega_{g})^{c_n}\mu_{g,n}}\ ,
\label{mummis}\ee
so that, since $\mu_{g,n}$ is nowhere vanishing and holomorphic on $\M_g$, it follows that even $[i_{N_n+1},\ldots,i_{M_n}|\tau]$ is holomorphic on $\M_g$
and vanishes when $\omega^{(n)}_{i_1},\ldots,\omega^{(n)}_{i_{N_n}}$ is
not a basis of $H^0(K^n_C)$. In particular, for $\tau$ belonging to the closure of the locus of hyperelliptic
Riemann period matrices $\mathcal{H}_g$ in ${\frak H}_g$, $[i_{N_n+1},\ldots,i_{M_n}|\tau]$ has zeroes of
order at least $(n-1)(g-1)-1$ \cite{Matone:2011ic}.

Since $\mu_{g,n}$ is invariant under the modular transformations
\be\label{modull}\tau
\mapsto \gamma\cdot\tau=(A\tau+B)(C\tau+D)^{-1}\ ,\ee
$\gamma\equiv\begin{pmatrix}A & B \\ C & D\end{pmatrix}\in \Gamma_g:={\rm Sp}(2g,\ZZ)$, it follows by (\ref{mummis}) that \cite{Matone:2011ic}
\begin{align}
\label{transfC}\sum_{j_{1},\ldots,j_{K_n}=1}^{M_n}\rho^{(n)}(\gamma)_{k_{1}j_{1}}\cdots\rho^{(n)}(\gamma)_{k_{K_n}j_{K_n}}&[j_{1},\ldots,j_{K_n}|\gamma\cdot\tau]\\
& = \det (C\tau+D)^{d_n} [k_{1},\ldots,k_{K_n}|\tau]\ ,\notag
\end{align}
where
$$\rho^{(n)}(\gamma)\cdot(\omega_{k_1}\cdots\omega_{k_n})= \sum_{j_1,\ldots,j_n=1}^g\omega_{j_1}\cdots\omega_{j_n}(C\tau+D)^{-1}_{j_1k_1}\cdots(C\tau+D)^{-1}_{j_nk_n}\ ,
$$
$k_1,\ldots,k_n=1,\ldots,g$, and
\be
d_n:=6n^2-6n+1-{g+n-1\choose n-1} \ .
\label{denned}\ee
Note that $d_2=g-12$,
so that $g=12$ is a critical point\footnote{It is worth noticing that $g=12$ is a critical value for the Ikeda lift and for the slope conjecture.}.

\section{Bosonic measure and the quadrics in $\PP^{g-1}$}

Interestingly, vector-valued modular forms are the natural generalization
of the classical elliptic modular forms, and therefore basic even for number theory, as seen by studying the cohomology
of the universal abelian variety. The theory of vector-valued modular forms is not particularly developed, only the case of low genus has been investigated. The fact that the Mumford forms define an infinite set of
vector-valued Teichm\"uller modular forms is of interest even in such a context.

A nice geometrical feature of the vector-valued Teichm\"uller modular forms derived from the Mumford forms is that these describe the Riemann surface (or canonical curve) in terms of quadrics in $\PP^{g-1}$.
Recently, such quadrics have been expressed as determinantal relations among quadratic differentials \cite{Matone:2006bb}.

An embedding of $C$ in $\PP^{g-1}$ is given by $p\mapsto
(\omega_1(p),\ldots,\omega_g(p))$. More generally, any element of $H^0(K_C^n)$
defines a homogeneous $n$-degree polynomial in $\PP^{g-1}$ by
$$
\phi^n:=\sum_{i_1,\ldots,i_n}B_{i_1,\ldots,i_n}\omega_{i_1}\cdots\omega_{i_n}\mapsto
\sum_{i_1,\ldots,i_n}B_{i_1,\ldots,i_n}X_{i_1}\cdots X_{i_n}\ ,$$
where $X_1,\ldots,X_g$ are homogeneous coordinates on $\PP^{g-1}$. A
basis of $H^0(K_C^n)$ corresponds to a basis of the homogeneous
polynomials of degree $n$ in $\PP^{g-1}$ that are not zero when
restricted to $C$. One may identify $C$ with the ideal of all
the polynomials in $\PP^{g-1}$ vanishing at $C$. Such an ideal
is generated by the quadrics
$$\sum_{j=1}^MC^i_jX^{(2)}_j=0\ ,\qquad\quad N+1\le i\le M\ ,$$
where we used the single index notation $X^{(2)}_j$ instead of $X_iX_j$. The canonical curves that are not
cut out by such quadrics are trigonal or isomorphic to a smooth plane
quintic. In these cases the ideal is
generated by the quadrics above together with a suitable set of
cubics corresponding to linear relations among
holomorphic $3$-differentials.

A property of the vector-valued Teichm\"uller modular forms
is that they provide the coefficients of such polynomials.
Namely, for each integer $n\geq 2$ and for all $i_{2},\ldots,i_{K_n}\in\{1,\ldots,M_n\}$ we have \cite{Matone:2011ic}
\be\sum_{i=1}^{M_n}[i,i_{2},\ldots,i_{K_n}|\tau]\omega^{(n)}_{i}(x)
=0\ .
\label{polyns}\ee In particular, for $n=2$ these are all the quadrics characterizing the canonical curve in projective space.
The explicit form of the coefficients of the quadrics has been derived in \cite{Matone:2006bb}.  However, finding their expression in terms of theta-constants is still an open question.
The vector-valued Teichm\"uller modular forms seem the natural candidates to find such expressions.

A related issue is the problem of the effective characterization of the Jacobian. This has been explicitly solved only for $g=4$:
there is a weight 8 Siegel modular form vanishing only on the Jacobian, this is the Schottky-Igusa form $F_4$.
We have seen that $F_4$ defines $[i|\tau]$. More generally, a similar relation between
vector-valued Teichm\"uller modular forms and the characterization of the Jacobian should hold even for $g>4$. To see this, let us first explicitly show the case $g=4$.
Let ${\cal I}_g$ be the closure of the locus of Riemann
period matrices in ${\frak H}_g$. Since $F_4$ vanishes
identically on $\mathcal{I}_4$, we have \cite{Matone:2011ic}
\be\label{Fgenquadr}0=\partial_t F_4(\tau)=\sum_{i\le j}{\partial
F_4\over
\partial Z_{ij}}\big\vert_{ Z=\tau}\partial_t\tau_{ij}=\sum_{i\le j}{\partial F_4\over
\partial Z_{ij}}\big\vert_{ Z=\tau}d\tau_{ij}(\partial_t)\ .\ee
It follows by (\ref{polyns}) that
$$
[ij|\tau]=c S_{4ij}(Z) \ ,
$$
$i,j=1,\ldots,g$, where $c$ is a constant. Note that here the double index notation $[ij|\tau]$ instead of
$[i|\tau]$, $i=1,\ldots,M$, has been used. It turns out that there is a simple relation relating the discriminant of the quadric and
$\chi_{68}$ \cite{Matone:2011ic}
$$\det S_4(\tau)=d\chi_{68}(\tau)^{1/2}\ , $$
where $d$ is a constant.
Note that $\det S_4$ and $\chi_{68}^{1/2}$ are modular forms (of weight
$34$) only when restricted to ${\cal I}_4$.
The above results also provide a straightforward proof that
for $g=4$, $n=2$, the Mumford form is
\be\mu_{4,2}=\pm{1\over c
S_{4ij}}{\omega_1\omega_1\wedge\cdots\wedge
\widehat{\omega_i\omega_j}\wedge\cdots\wedge \omega_4\omega_4\over
(\omega_1\wedge\cdots\wedge\omega_4)^{13}} \ ,
\label{gquattro}\ee
as conjectured by Belavin and Knizhnik
\cite{Belavin:1986cy} and by Morozov \cite{MorozovDA}.

\section{Vector-valued Teichm\"uller modular forms as determinants}

A basic question is to understand what is the analog structure of $\mu_{4,2}$ in the case of $\mu_{g,n}$, for arbitrary $g$ and $n$. Are there forms that, like $F_4$ generate the
vector-valued Teichm\"uller modular forms for any $n$ and $g$? This may shed light on algebraic-geometrical structures underlying
the bosonic and supersymmetric string measures.

There is a natural answer to the above question.
\begin{theorem}\label{Thuno}
\be\label{cij}
[i_{N_n+1},\ldots,i_{M_n}|\tau]= \det_{\substack{j\\
k\in\{ {N_n+1},\ldots,{M_n}\}}} C_{i_k}^j(\tau) \ ,
\ee
where $C^j(\tau)\equiv(C_1^j(\tau),\ldots,C_{M_n}^j(\tau))$ are $K_n=M_n-N_n$ linearly independent vectors, non-trivial solutions of the degree $n$ polynomial equations
\be
\sum_{i=1}^{M_n}C_{i}^j(\tau)\omega_i^{(n)}=0 \ ,
\label{richiamo}\ee
$j=1,\ldots,K_n$. Furthermore,
\be
\sum_{i=1}^{M_n}\rho^{(n)}(\gamma)_{ki} C_i^j(\gamma\cdot\tau)=\sum_{l=1}^{K_n}A_l^j\det(C\tau+D)^{n_l} C_k^l(\tau) \ ,
\label{trC}\ee
$k=1,\ldots,M_n$, where $\det A=1$ and
\be
\sum_{k=1}^{K_n} n_k=d_n \ .
\label{questa}\ee
\end{theorem}

\vskip 6pt

\noindent {\sl Proof.}
To prove Eq.(\ref{cij}) let use the fact that there are $K_n$ linearly independent vectors $D^j(\tau)\equiv(D_1^j(\tau),\ldots,D_{M_n}^j(\tau))$ non-trivial solutions of
\be
\sum_{i=1}^{M_n}D_{i}^j(\tau)\omega_i^{(n)}=0 \ ,
\label{richiamobbb}\ee
$j=1,\ldots,K_n$. Then we rewrite this in the form
$$
\sum_{k\in M_n\backslash\{1,\ldots,N_n-1,N_n+1\}}D_{i_k}^j(\tau)\omega_{i_k}^{(n)}=-\sum_{k\in\{1,\ldots,N_n-1,N_n+1\}}D_{i_k}^j(\tau)\omega_{i_k}^{(n)} \ ,
$$
to express $\omega_{i_{N_n}}$ in
$$
\omega_{i_1}^{(n)}\wedge\cdots \wedge\omega_{i_{N_n}}^{(n)} \ ,
$$
in terms of $\omega_{i_1}^{(n)},\ldots,\omega_{i_{N_n}-1}^{(n)},\omega_{i_{N_n}+1}^{(n)}$, to find that, up to a sign,
$$\det_{\substack{j\\
k\in\{{N_n},{N_n+2},\ldots,{M_n}\}}}D_{i_k}^j(\tau)\omega_{i_1}^{(n)}\wedge\cdots \wedge\omega_{i_{N_n}}^{(n)} \ ,
$$ is equal to
$$
\det_{\substack{j\\
k\in\{{N_n+1},\ldots,{M_n}\}}} D_{i_k}^j(\tau)
\omega_{i_1}^{(n)}\wedge\cdots \wedge\omega_{i_{N_n}-1}^{(n)}\wedge\omega_{i_{N_n}+1}^{(n)} \ .
$$
Since $$\det_{\substack{j\\
k\in\{{N_n+1},\ldots,{M_n}\}}} D_{i_k}^j(\tau) \ ,$$ satisfies Eq.(\ref{polyns}) and is completely antisymmetric as $[i_{N_n+1},\ldots,i_{M_n}|\tau]$, it follows that they differ by a global term given by the determinant of the non degenerate
matrix $X$ such that
$C^j=\sum_{l=1}^{K_n}X_l^jD^l$. Eqs.(\ref{trC}) and (\ref{questa}) follow
by (\ref{transfC}) and (\ref{denned}) respectively. \hfill$\square$

\section{The Mumford-Polyakov form at any genus}

A consequence of Theorem \ref{Thuno} is that the bosonic measure, corresponding to the case $n=2$, is built in terms of the determinant of the coefficients of the quadrics describing the
Riemann surface in $\PP^{g-1}$. Also, note that the range of the indices of the determinant is from 1 to $K=M-N=(g-2)(g-3)/2$, which is the codimension of the locus of Riemann period matrices in the
Siegel upper half-space. This is another piece of evidence that the Schottky problem is strictly related to string theory.

Let us choose an
arbitrary point $p_0\in C$ and
consider the Abel-Jacobi
map $I(p):=(I_1(p),\ldots,I_g(p))$
$$I_i(p):=\int_{p_0}^p\omega_i\ ,$$ $p\in C$. It embeds $C$ into the Jacobian $J(C):={\CC}^g/L_\tau$,
$L_\tau:={\ZZ}^g +\tau {\ZZ}^g$.

Denote by $\Theta_{s}$ the locus of the $e\in \ZZ^g$ where $\theta(e,Z)$ and its gradient vanish.
It turns out that
the map $P\mapsto\theta(I(P)\pm e)$, $e\in J(C)$ vanishes identically on
$C$ for all $P_0\in C$ if and only if $e\in\Theta_s$. As a consequence ${\partial^2\theta\over\partial u_j\partial u_k}(\pm e)$ and ${\partial^3\theta\over\partial u_j\partial u_k\partial u_l}(\pm e)$
are the coefficients of the quadrics and cubics describing $C$ in $\PP^{g-1}$, that is
\be
\sum_{j,k=1}^g{\partial^2\theta\over\partial u_j\partial u_k}(\pm e)\omega_j\omega_k=0 \ ,
\label{quaa}\ee
\be
\sum_{j,k,l=1}^g{\partial^3\theta\over\partial u_j\partial u_k\partial u_l}(\pm e)\omega_j\omega_k\omega_l=0 \ .
\label{cubb}\ee
By means of the heat equation
$$
{\partial^2\theta(u,Z)\over\partial u_j\partial u_k}=2\pi i(1+\delta_{jk}){\partial \theta(u,Z)\over\partial Z_{jk}} \ ,
$$
and using the single indexing, one may rewrite Eq.(\ref{quaa}) in the form
\be
\sum_{j=1}^K{\partial\theta\over\partial Z_j}(\pm e)|_{Z=\tau}\omega^{(2)}_j=0 \ .
\label{quaadd}\ee
Since as $e$ varies in $\Theta_s$ Eq.(\ref{quaa}) generates the ideal of quadrics passing through the
curve $C$, it follows by (\ref{richiamo}) that there are $e^1,\ldots,e^K\in\Theta_s$, such that the $C_i^j$ in Theorem \ref{Thuno} are
$$
C^j_i(\tau)=\sum_{l=1}^K\Lambda^{j}_l{\partial\theta\over\partial Z_i}(e^l)|_{Z=\tau} \ ,
$$
$j=1,\ldots,K$, for some matrix $\Lambda^{j}_k$, so that
$$
[i_{N+1},\ldots,i_{M}|\tau]= \det_{\substack{j\\
k\in\{ {N+1},\ldots,{M}\}}}\sum_{l=1}^K\Lambda^{j}_l{\partial\theta\over\partial Z_{i_k}}(e^l)|_{Z=\tau} \ .
$$
Since $\theta(e^l)$ vanishes for $Z=\tau$, we have the basic fact
that the vector-valued Teichm\"uller modular forms are given by a
Jacobian. Set \be
F_g^j(\tau):=\sum_{l=1}^K\Lambda^{j}_l\theta(e^l)|_{Z=\tau} \ .
\label{eccoquas}\ee
We have proved the following
\begin{theorem}\label{Thdue}
\be
[i_{N+1},\ldots,i_{M}|\tau]= \det_{\substack{j\\
k\in\{ {N+1},\ldots,{M}\}}}{\partial F_g^j(Z)\over\partial Z_{i_k}}|_{Z=\tau} \ .
\label{cijtre}\ee
\end{theorem}

\noindent As a corollary, it follows by (\ref{mummis}) that for any $g$ the Mumford-Polyakov form is
\be
\mu_{g,2}={\epsilon_{i_1,\ldots,i_M} \omega_{i_1}^{(2)}\wedge\cdots\wedge\omega_{i_N}^{(2)}\over(\omega_1\wedge\cdots\wedge\omega_g)^{13}\det_{\substack{j\\
k\in\{ {N+1},\ldots,{M}\}}}{\partial F_g^j(Z)\over\partial Z_{i_k}}|_{Z=\tau}} \ .
\label{okkapissimo}\ee
Such an expression for $\mu_{g,2}$ provides strong evidence that the Mumford-Polyakov form can be expressed as a multiresidue. More precisely, if the unimodular matrix $A$ in Theorem \ref{Thuno}
is diagonal and
under suitable conditions on the divisors of $F_g^j(Z)$,
the bosonic partition function could be then expressed as a multiresidue on the Siegel upper half-space
\be\label{wonderfulformula}
Z_g=\int_{{\frak H}_g} {1\over\det({\rm Im} Z)^{13}} \Bigg|{\prod_{i\leq j}^{g}dZ_{ij}\over\prod_{j=1}^{(g-2)(g-3)/2} F_g^j(Z)}\Bigg|^2 \ .
\ee
If the above properties for $F_g^j(Z)$ could be satisfied, then finding them would provide an effective solution of the Schottky problem.
In the case of $g=5$, it is known that the Jacobian variety $J(C)$ is not characterized by the vanishing of three globally defined Siegel modular forms. Nevertheless,
by (\ref{trC}) it follows that the weights $n_j$ of $F_g^j(Z)$ must satisfy the condition
$$
\sum_{j=1}^K n_j=12-g \ ,
$$
which implies that in general $F_g^j(Z)$ are not holomorphic on ${\frak H}_g$. In particular, note that the $n_j$ may be not integers.
If the unimodular matrix $A$ in (\ref{trC}) is not diagonal, then one should investigate
a different expression of the integrand in (\ref{wonderfulformula}) and construct invariants from $F_g^j$ and $\bar F_g^j$. In this respect, it would be interesting to check
possible non homogeneous terms in their transformation properties and/or a dependence of $F_g^j$ on $\bar Z$ that either vanishes on $J(C)$ or that disappears once one considers
its Jacobian on $J(C)$. It may be that some higher genus generalization of the Mock theta functions play a r$\rm\hat o$le in such a context.

The construction in such a section can be extended to Mumford forms with arbitrary $n$. This is related to the generalized Beltrami differentials associated
to $H^0(K^n_C)$ \cite{Matone:1993tj}. In particular, one should extend the map $\omega_i^{(2)}\mapsto{1\over2\pi i}d\tau_i^{(2)}$ to arbitrary $n$
$$\omega_i^{(n)}\mapsto{1\over2\pi i}d\tau_i^{(n)} \ ,
$$
that defines the tangent space to the moduli space associated to the
holomorphic $n$-differentials, that is the moduli space of vector
bundles on Riemann surfaces. This is related to the chiral split for
the higher order diffeomorphism anomalies. The Wess-Zumino
conditions correspond to the cocycle identities (see section 3.4 of
\cite{Matone:1993tj}).

We observe that recently the problem of superstring perturbation theory attracted renewed interest \cite{D'Hoker:2001zp}-\cite{DuninBarkowski:2012ya}.
Interestingly, the vector-valued Teichm\"uller modular forms also appear in the superstring measure, in particular the cosmological constant for $g=4$ turns out to be
given in terms of $\mu_{4,2}$ and $F_4$, namely
\be
{F_4\over
\partial_{\tau_j} F_4}{d\tau_1\wedge\cdots\wedge
\widehat{d\tau_j}\wedge\cdots\wedge d\tau_{44}\over
(\omega_1\wedge\cdots\wedge\omega_4)^{5}}=0 \ ,
\label{gquattrobisse}\ee
and vanishes just because $F_4$ characterizes the Jacobian locus. $F_g$ gives a vanishing cosmological constant also for $g\leq3$.
The fact that $F_g$ is the difference between of the two theta series associated to the even unimodular lattices $E_8$ and $D_{16}^+$
of the two heterotic strings may be due to foundational aspects. It would be natural to guess that $F_g$ characterizes the cosmological constant for any $g$.
However, it has been shown in \cite{Matone:2010yv} that getting
the amplitudes by degenerating the Riemann surfaces, one sees that after GSO projection, the two-point function does not vanish at genus four,
as expected from space-time supersymmetry
arguments. This is a strong hint that the Grushevsky ansatz \cite{Grushevsky:2008zm}
the superstring measure should be corrected for $g>4$. Very recently in an interesting paper,
Dunin-Barkowski, Sleptsov and Stern \cite{DuninBarkowski:2012ya} have shown that there is a minor modification of Grushevsky ansatz \cite{Grushevsky:2008zm} leading to a vanishing cosmological constant even for $g=5$ that reproduces, by degeneration, a vanishing two-point function at $g=4$. However, they also proved that a similar mechanism fails starting from $g=6$.

\vspace{.6cm}

\noindent
Let us mention some possible developments in investigating the superstring measure.

\begin{enumerate}

\item {\it Focusing on the geometry of the Jacobian locus}. Starting from $g=5$ there are more forms, namely $K$ forms, vanishing on the Jacobian and that define the bosonic measure. If the structure of the cosmological constant at any genus should reproduce the one of $g=4$, factorization arguments suggest that the $F_g^j$, $j=1,\ldots,K$, should appear in the cosmological constant too, not only in the bosonic part of the measure. This is related to the fact that there are not stable Schottky forms, as recently shown by
 Codogni and Shepherd-Barron \cite{CS}.

\item {\it Trigonal curves}. Another possibility is to focus on the curious ubiquity of $F_g$ in the expression for the cosmological constant up to $g=4$. It is known that for $g=5$ it vanishes only on the locus of $\M_g$ corresponding to trigonal curves. This means that if one restricts the path-integral formulation to trigonal curves, then the Grushevsky ansatz would be effective at least up to $g=5$. In this respect, it is interesting to note that trigonal curves, related to triality, have, with their three $\PP^1$, a structure reminiscent of the geometrical object defining the interactions in string theory, the pant. In this context it may be useful to recall that at the beginning of the superstring era several works were dedicated to the check, in perturbation theory, of the unitarity constraint.

\item {\it $n(g)$-gonal curve}. A property of $F_4$ is that it identifies a subvariety, the Jacobian, which is of codimension 1. For $g>4$ the codimension of $J(C)$ is $K$, so that there is not a divisor vanishing only on $J(C)$. This may be a problem in focusing on the trigonal curves. A possibility is that the case of $g=4$ is the first one of a more subtle mechanism.\footnote{I thank A. Sagnotti and R. Volpato for enlighting suggestions on the following analysis.} Actually, one should consider not just trigonal curves but $n(g)$-gonal curves. In particular, one should consider $n(g)$-gonal curves with $n(g)$ fixed by requiring that the codimension of their moduli spaces in $\M_g$ be the lowest one.
        A $n(g)$-gonal curve of genus $g$ is a $n$-sheet covering of $\PP^1$, so that, by Hurwitz theorem, it should have $2g+2n-2$ ramification points. Fixing three points by the ${\rm PSL}(2,\CC)$ symmetry of $\PP^1$, one sees that
        the locus of the moduli space of $n(g)$-gonal curves in $\M_g$ is $2g+2n-5$. Requiring that the $n(g)$-gonal locus be of codimension one in $\M_g$ fixes
        $$
        2n(g)=g+1 \ .
        $$
        It follows that for $g$ odd the space of $(g+1)/2$-gonal curves is a divisor in $\M_g$. For $g$ even, all the curves are $(g+2)/2$-gonals, with the moduli space of $g/2$-gonal curves being of codimension 2 in $\M_g$. It follows that a possible generalization of $F_4$ is to consider a form such that its zero divisor in $\M_g$ be the locus of the $(g+1)/2$-gonal curves for $g$ odd, and properly containing the moduli of the $g/2$-gonal curves for $g$ even.
        Another possibility is that such a form always contains the trigonal locus, but for $g>5$ only as sublocus. In this case, it would be interesting to understand if there exists a modular form which is a Schottky form for $g=6$ and defines, by its zeroes, the divisor of the 4-gonals at $g=7$.

\end{enumerate}

The fact that $g=5$ is a critical value for the superstring measure also follows by a recent result by Donagi and Witten \cite{D}\cite{W}. They proved that at least for $g\geq 5$ the supermoduli space is not a split
supermanifold, that is the moduli space of super Riemann surfaces does not map to the moduli space of Riemann surfaces with a spin structure. In this context, it should be observed that the
 Berkovits pure spinor formulation of superstring theory is defined on Riemann surfaces \cite{BerkovitsFE}\cite{geometrica}. Understanding the relation between GSO projection in the RNS formalism and the Berkovits pure spinor formulation, where the GSO projection is absent, may lead to understand whether the Donagi-Witten result forbids a priori the reduction to $\M_g$.

Finally, we note that Eq.(\ref{wonderfulformula}) suggests that the bosonic string may be formulated on the Siegel upper half-space. Perhaps one may use the properties of the period matrices associated to the Prym varietes. An investigation in this direction may also lead to find a super analog of (\ref{wonderfulformula}) on the super Siegel half-space.

\section*{Acknowledgements}

I thank  G. Codogni, R. Donagi, S. Grushevsky, Yu. Manin, M. Porrati, A. Sagnotti, R. Salvati Manni, M. Tonin, R. Volpato, A. Voronov and E. Witten for helpful discussions/correspondence.
This research is supported by the Padova
University Project CPDA119349 and by the MIUR-PRIN contract 2009-KHZKRX.

\end{document}